# Efficiency analysis for quantitative MRI of T1 and T2 relaxometry methods

**Authors:** David Leitão[1], Rui Pedro A. G. Teixeira[1,2], Anthony Price[1,2], Alena Uus[1], Joseph V. Hajnal[1,2] and Shaihan J. Malik[1,2]

[1]School of Biomedical Engineering and Imaging Sciences, King's College London, London, United Kingdom
[2]Centre for the Developing Brain, King's College London, London, United Kingdom

# Abstract

This study presents a comparison of quantitative MRI methods based on an efficiency metric that quantifies their intrinsic ability to extract information about tissue parameters. Under a regime of unbiased parameter estimates, an intrinsic efficiency metric $\eta$ was derived for fully-sampled experiments which can be used to both optimize and compare sequences. Here we optimize and compare several steady-state and transient gradient-echo based qMRI methods, such as magnetic resonance fingerprinting (MRF), for joint T1 and T2 mapping. The impact of undersampling was also evaluated, assuming incoherent aliasing that is treated as noise by parameter estimation. In-vivo validation of the efficiency metric was also performed.

Transient methods such as MRF can be up to 3.5 times more efficient than steady-state methods, when spatial undersampling is ignored. If incoherent aliasing is treated as noise during least-squares parameter estimation, the efficiency is reduced in proportion to the SNR of the data, with reduction factors of 5 often seen for practical SNR levels. In-vivo validation showed a very good agreement between the theoretical and experimentally predicted efficiency. This work presents and validates an efficiency metric to optimize and compare the performance of qMRI methods. Transient methods were found to be intrinsically more efficient than steady-state methods, however the effect of spatial undersampling can significantly erode this advantage.

# Introduction

Over time many methods have been developed that aim to estimate $T_1$ and $T_2$ as effectively as possible, from classical inversion-recovery and spin-echo sequences, to steady-state sequences (Deoni, Rutt and Peters, 2003; Welsch *et al.*, 2009; Heule, Ganter and Bieri, 2014; Teixeira, Malik and Hajnal, 2017; Shcherbakova *et al.*, 2018) and more recently MR Fingerprinting (MRF) (Ma *et al.*, 2013). Selecting which method to favor for any given scenario can be challenging since the achieved precision (statistical uncertainty) and accuracy (proximity to true value), and the way these change with acquisition time are generally complex functions of the pulse sequence settings, tissue properties, specific details of the hardware used and the type of image reconstruction.

Nevertheless, comparisons have been made using metrics that strive to evaluate the intrinsic merits of each method (Crawley and Henkelman, 1988; Jones *et al.*, 1996; Deoni, Rutt and Peters, 2003; Ma *et al.*, 2013; Assländer, Novikov, *et al.*, 2018; Assländer, 2020), accounting for the differences external to the methods themselves. Fundamentally, these differences are the amount of data and the SNR of the experiment. The amount of data has been normalized using either the total number of measurements (Jones *et al.*, 1996) or the total acquisition time (Edelstein *et al.*, 1983; O'Donnell, Gore and Adams, 1986; Crawley and Henkelman, 1988; Deoni, Rutt and Peters, 2003; Ma *et al.*, 2013; van Valenberg *et al.*, 2017; Assländer, 2020). The SNR of the experiment has been normalized using the voxel volume (Deoni, Rutt and Peters, 2003), a combination of the thermal noise level with the proton density (Edelstein *et al.*, 1983; O'Donnell, Gore and Adams, 1986; Jones *et al.*, 1996; van Valenberg *et al.*, 2017; Assländer, Novikov, *et al.*, 2018; Assländer, 2020) or the signal dynamic range of each method (Crawley and Henkelman, 1988). Some of these studies (Ma *et al.*, 2013; Assländer, Novikov, *et al.*, 2018; Assländer, 2020) showed that balanced MRF (using a balanced readout) outperforms Driven Equilibrium Single Pulse Observation of $T_1/T_2$ (DESPOT) for both $T_1$ and $T_2$ estimation, but it is as yet unclear how other commonly used methods, like spoiled MRF (Jiang *et al.*, 2015) (using a gradient spoiled readout) or Double Echo Steady State (DESS) (Welsch *et al.*, 2009), compare. Furthermore, methods based entirely on gradient spoiled readouts are popular for their insensitivity to off-resonance at a cost of an SNR penalty – it is unclear whether this trade-off benefits $T_1$ and $T_2$ estimation.

In this work we focus on precision and propose a general efficiency metric that integrates the concepts from figures of merit used so far. The resulting metric is then used to optimize and comprehensively compare a range of well-established methods that simultaneously estimate $T_1$ and $T_2$. Finally, we discuss the utility of the efficiency metric and other important considerations when comparing qMRI methods.

# Theory

We consider that each voxel contains a single pool of spins characterized by unique values of $T_1$ and $T_2$; that the signal models for the different qMRI methods are accurate, subject to additive Gaussian noise; and that parameter estimation results in an unbiased estimate of the parameters of interest $\theta$ (i.e. $T_1$ and $T_2$ but potentially other parameters). Therefore, the error in the estimates is defined by the precision that is characterized by the standard deviation $\sigma_\theta$ of the estimated parameter $\theta$. Analogous to the definition of SNR, the precision can also be represented by the parameter-to-noise ratio ($\theta NR$):

$$\theta NR = \frac{\theta}{\sigma_\theta} \qquad [1]$$

Although the $\theta NR$ directly relates to the SNR, it also depends on how much information about the parameter being measured is encoded in the data. Further, the SNR can be broken down to consist of an intrinsic SNR (relating to the receiver system, field strength, resolution etc.) and the amount of data acquired. These dependencies have been highlighted in previous works (Edelstein *et al.*, 1983; O'Donnell, Gore and Adams, 1986; Crawley and Henkelman, 1988; Jones *et al.*, 1996) and can be expressed in the following equation that serves to define efficiency with which a parameter $\theta$ is estimated, $\eta(\theta)$:

$$\theta NR = \eta(\theta) \cdot SNR_{max} \cdot \sqrt{T_{acq}} \qquad [2]$$

Here, $SNR_{max} \equiv M_0/\sigma_0$ represents the maximum SNR of any one measurement. $M_0$ is defined as the maximum signal from a voxel that would be measured by applying a 90° pulse with the magnetization in thermal equilibrium; this is a characteristic of the system (field strength and coil) and acquisition geometrical parameters (resolution and field of view); $\sigma_0$ is the receiver noise standard deviation (i.e., what would be measured during one k-space data readout scaled to account for differences in scaling between k-space and image domain) and is characteristic of the k-space readout and its bandwidth; $T_{acq}$ is the total acquisition duration for all data required to estimate the parameter $\theta$. Rewriting Eq. [2] with $\eta(\theta)$ as its subject gives:

$$\eta(\theta) = \frac{\theta NR}{SNR_{max}} \frac{1}{\sqrt{T_{acq}}} = \frac{\theta}{\sigma_\theta} \frac{\sigma_0}{M_0} \frac{1}{\sqrt{T_{acq}}} \leq \frac{\theta}{\sigma_\theta^{CRLB}} \frac{\sigma_0}{M_0} \frac{1}{\sqrt{T_{acq}}} \qquad [3]$$

An upper bound on the efficiency can be obtained without need for physical measurement by calculating the Cramér-Rao Lower Bound (CRLB) for $\sigma_\theta$ ($\sigma_\theta^{CRLB}$)(Sengupta and Kay, 1995), resulting in $\eta(\theta) \leq \eta^{CRLB}(\theta)$. The ability to achieve this bound depends on the full parameter reconstruction pipeline, which we assume to extract all encoded information so that equality holds.

To illustrate the utility of the efficiency concept, Figure 1 shows an example simplified 'fingerprinting' experiment with only 5 radiofrequency (RF) pulses, each followed by a measurement. When optimized for maximum efficiency in estimating $T_1$ and $T_2$, two pulses get set to zero amplitude leaving a 3 pulse, 60°-180°-90°, sequence that basically combines two familiar sequences: a spin-echo and an inversion recovery with optimized inversion and echo times for maximum information. Interestingly, it is more efficient to measure *fewer* signals but allow magnetization to recover, in this case.

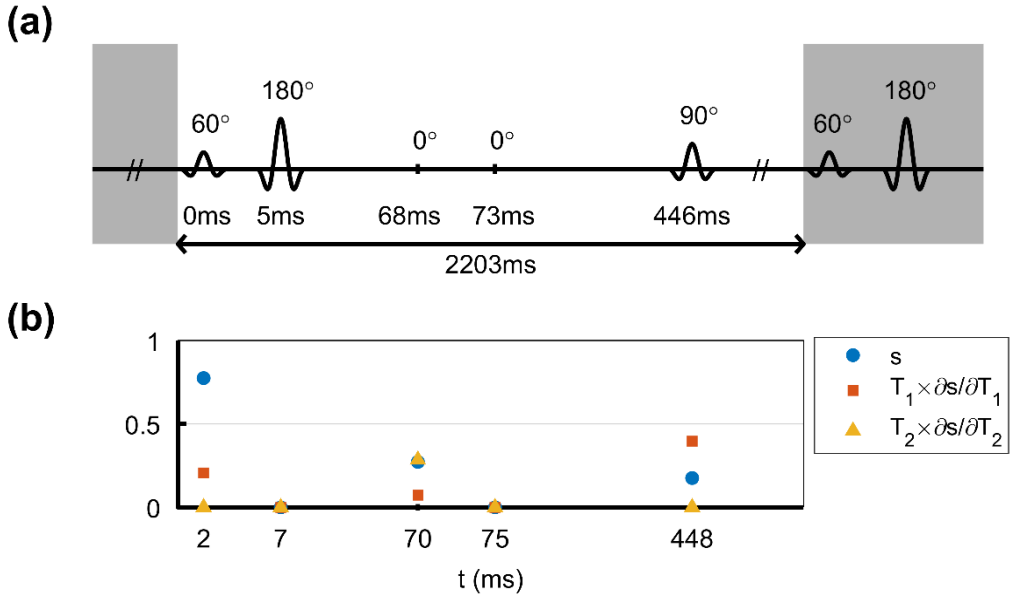

Figure 1: Optimized fingerprint with just 5 pulses that are applied cyclically, with a spoiling gradient preceding each pulse. Schematic representation of (a) the optimized flip angles and intervals between the pulses and (b) the signal $s$ and its derivatives w.r.t. $T_1$ and $T_2$ at echo time (2ms). Note that the regions in gray are already repetitions of the main block of 5 pulses. Because the efficiency measure incorporates the acquisition time it allows the best structure of flip angles and their timings to be found such its averaging extracts the most information about the parameters of interest.

Although the results from Eqs. [2-3] are general, when computing efficiency with the CRLB we have assumed that each measurement is fully-sampled and acquired using a single channel coil with uniform sensitivity. This allows a quick evaluation of efficiency, crucial for use in optimization, but is not realistic. In practice, data is acquired using multi-channel receiver coils, often with some degree of undersampling, either for parallel imaging or as an inherent part of the method as in MRF. For multi-channel coils, $\sigma_0$ is still defined as the standard deviation in one readout in one channel, as after pre-whitening all channels should have the same noise level (Pruessmann et al., 1999). In this case $M_0$ is the result obtained from optimally combining all channels. Use of multi-channel coils with optimal combination would not affect the efficiency; better intrinsic SNR would improve the $\theta NR$ which is captured by $SNR_{max}$ in Eq. [3]. On the other hand, undersampling will affect both $T_{acq}$ and $\sigma_\theta$, impacting the efficiency of the sequence. According to Hu and Peters (Hu and Peters, 2019), the standard deviation ($\sigma_{\theta,R}$) of $\theta$ in an experiment with an $R$ undersampling factor can be related to the fully-sampled case ($\sigma_\theta$) as:

$$\frac{\sigma_{\theta_n,R}}{\sigma_{\theta_n}} = d_R(\theta_n)\sqrt{R} \qquad [4]$$

where $d_R$ is the so-called "dynamics-factor" that expresses the parameter error amplification due to the ill-conditioning of the parameter estimation (Hu and Peters, 2019). The subscript $n$ represents the $n^{th}$ voxel, as $d_R$ may be spatially varying since it will include effects from coil encoding such as the g-factor (Pruessmann *et al.*, 1999) as well as sampling effects that may arise from use of time-varying non-cartesian k-space trajectories. Hence the efficiency of an undersampled experiment $\eta_R$ can be related back to the fully-sampled experiment efficiency $\eta$ by:

$$\eta_R(\theta_n) = \frac{\eta(\theta_n)}{d_R(\theta_n)} \qquad [5]$$

For a least-squares estimator $d_R \geq 1$, so can only reduce the efficiency compared to the fully-sampled experiment. In general, $d_R$ must be estimated by analysis of the full parameter reconstruction pipeline, which will be problem dependant and could become very large for a non-cartesian image reconstruction. In the following subsection we approximate $d_R$ for the special case of a zero-filled reconstruction.

Zero-filled reconstruction

The original MRF paper (Ma *et al.*, 2013) proposed a zero-filled reconstruction that treated undersampling artifacts as noise. In this case for a least-squares estimator the final parameter standard deviations are proportional to the signal standard deviations; thermal noise and 'aliasing noise' both contribute noise that have a similar impact on parameter estimation but may differ in relative strength. Assuming 'aliasing noise' follows an independent Gaussian distribution $N(0, \sigma_{alias}^2)$, the image-domain signal standard deviation in an undersampled experiment $\sigma_{image,R}$ may be written:

$$\sigma_{image,R} = \sqrt{\sigma_{alias}^2 + R \cdot \sigma_{image}^2} \qquad [6]$$

Where $\sigma_{image}$ is the image-domain noise standard deviation in a fully-sampled experiment. Hence, we may write

$$d_R \approx \frac{1}{\sqrt{R}} \frac{\sigma_{image,R}}{\sigma_{image}} = \sqrt{\frac{1}{R} \frac{\sigma_{alias}^2}{\sigma_{image}^2} + 1} \qquad [7]$$

This may further be written in terms of the signal-to-noise ratio in a 'fully-encoded' image ($SNR_{image}$) and the signal-to-aliasing ratio ($SaR_{image}$) from an undersampled image without thermal noise as:

$$d_R \approx \sqrt{\frac{1}{R}\frac{SNR_{image}^2}{SaR_{image}^2} + 1} \qquad [8]$$

It is therefore expected that the dynamics factor will become larger if the signal-to-noise ratio of the data improves; although at first this seems non-intuitive it highlights that aliasing effects are proportional to the signal, so a stronger signal leads to a larger contribution. Note that in Eqs.[6-8] we have considered a constant signal amplitude and an image domain SNR since aliasing is fundamentally treated in the image domain.

# Method

### Optimized sequence design

In order to make a fair comparison between all analyzed methods, sequence acquisition settings $\boldsymbol{u}$ (repeat time, flip angles ($\alpha$) etc.) were optimized similarly to other works (Gras *et al.*, 2017; Nataraj, Nielsen and Fessler, 2017; Teixeira, Malik and Hajnal, 2017; Assländer *et al.*, 2019; Zhao *et al.*, 2019), improving $T_1$ and $T_2$ efficiencies for a range of tissue parameters $\boldsymbol{p}$ represented by the set of parameters $P$. For clarity, $\theta$ is the set of parameters of interest, while $\boldsymbol{p}$ is all parameters that may affect the signal; in general $\eta(\theta) \equiv \eta(\theta; \boldsymbol{p}, \boldsymbol{u})$ but this dependence is kept implicit for notation simplicity. In this work $\theta = \{T_1, T_2\}$ while $\boldsymbol{p}$ depends on the types of sequence used – if spoiled sequences are used then $\boldsymbol{p} = \{T_1, T_2, M_0\}$, but if balanced sequences are used, then off-resonance frequency $\omega_0$ and phase of the measurement $\phi_0$ are also relevant, such that $\boldsymbol{p} = \{T_1, T_2, M_0, \phi_0, \omega_0\}$.

The optimization solved to find the acquisition parameters $\boldsymbol{u}$ that maximizes efficiency for each method follows:

$$\boldsymbol{u}^{opt} = \arg\min_{\boldsymbol{u}} \sum_{\boldsymbol{p} \in P} \left(\frac{1}{\eta(T_1; \boldsymbol{p}, \boldsymbol{u})}\right)^2 + \left(\frac{1}{\eta(T_2; \boldsymbol{p}, \boldsymbol{u})}\right)^2 \qquad [9]$$
$$\text{s.t.} \quad g(\boldsymbol{u}) \leq 0$$
$$\quad f(\boldsymbol{u}) = 0$$

where $g$ and $f$ are method-dependent constraint functions and are detailed in Supporting Information Table S1. The set of parameters $P$ consists of $T_1 = 781$ms, $T_2 = 65$ms, (corresponding to white matter at 3T (Bojorquez *et al.*, 2017)); $M_0 = 1$, $\phi_0 = 0°$ and $\omega_0 \in [-100, 100]$Hz in steps of 5Hz. Inclusion of a range of $\omega_0$ forces methods based on balanced sequences to achieve good efficiencies over a range of frequencies.

The optimization problem was solved using the Sequential Quadratic Program algorithm from Matlab function *fmincon*. Whenever the number of design variables was $\leq 400$, a multi-start strategy was employed consisting of 100 random initializations, otherwise a single initialization was used consisting of the originally published acquisition settings for the respective method.

## Efficiency comparison

We have studied five steady-state methods: DESPOT (Deoni, Rutt and Peters, 2003) and a variant called Joint System Relaxometry (JSR) (Teixeira, Malik and Hajnal, 2017) (analyzed together due to their similarity), PLANET (Shcherbakova *et al.*, 2018), DESS (Welsch *et al.*, 2009) and Triple Echo Steady State (TESS) (Heule, Ganter and Bieri, 2014); and MRF sequences with gradient spoiled (Jiang *et al.*, 2015) or balanced (Ma *et al.*, 2013) readouts. MRF sequences started from thermal equilibrium or were in a Driven Equilibrium (DE) mode (Ma *et al.*, 2018; Assländer *et al.*, 2019), in which a pulse train of fixed length is cycled such the final magnetization is equal to the initial magnetization. For each method optimized acquisition settings $\boldsymbol{u}^{opt}$ were determined using Eq.[9] with efficiency calculated using the CRLB in Eq.[3]. This was repeated for several acquisitions with different numbers of measurements (Table 1). The signal model, optimization constraints and acquisition settings for each method are in Supporting Information Table S1.

Table 1: Number of measurements of the several acquisitions for which the $T_1$ and $T_2$ efficiencies of every method were optimized. $N$ is the number of measurements in the transient method (length of the fingerprint). For the transient methods (in orange), fingerprints with less than 400 measurements (in gray) were not considered for further analysis as these could be incompatible with spatial encoding.

| Method | | Number of measurements |
|---|---|---|
| DESPOT/JSR | | All feasible combinations of SPGR and bSSFP measurements from a minimum 3 measurements up to a maximum of 8; |
| PLANET | | From 3 bSSFP measurements up to 20 bSSFP measurements; |
| DESS | | From 2 DESS measurements up to 8 DESS measurements; |
| TESS | | From 1 TESS measurement up to 6 TESS measurements; |
| Non-Driven Equilibrium | Spoiled and Balanced MRF | $N = \{5, 10, 20, 50, 100, 200, 300, 400, 500, 600, 700, 800, 900, 1000, 1200, 1400, 1600\}$ |
| Driven Equilibrium | Spoiled and Balanced MRF | $N = \{5, 10, 20, 50, 100, 200, 300, 400, 500, 600, 700, 800, 900, 1000\}$ |

For each method the acquisition with the lowest cost function (highest efficiency) was evaluated for a wider range of $T_1 \in [600, 1200]\,ms$ in steps of $40\,ms$ and $T_2 \in [40, 100]\,ms$ in steps of $4\,ms$. We found that MRF sequences with very small numbers of pulses can achieve very high efficiencies, particularly if starting from thermal equilibrium. However, in practice such short pulse trains make only a limited number of measurements, so cannot support spatial encoding. Therefore, MRF acquisitions with less than 400 excitations were not evaluated over the extended parameter range (values shown in gray in Table 1).

For the main cross comparisons between methods, all efficiencies were calculated assuming fully-sampled measurements. This makes a concise and transparent presentation and is a reasonable approach for steady-state methods. However, MRF methods are most often acquired with a considerable degree of undersampling, and in these cases the obtained efficiency would also depend on the dynamics factor $d_R$ according to Eq.[5]. To address this we estimated $d_R$ for the sub-optimal zero filled reconstruction using Eq.[8]. Both random and spiral undersampling were explored for the Shepp-Logan phantom and Monte-Carlo simulations (100,000 trials each with independent Gaussian additive noise) were performed to estimate the standard deviation of the undersampled data $\sigma_{image,R}$. Several undersampling factors $R$ and different SNR levels were used to estimate $d_R$.

Validation experiments

To experimentally validate the efficiency metric the standard deviation of $\theta$ needs to be determined, requiring multiple estimates obtained from different datasets. To stay within an acceptable acquisition time, we focused on validation of efficiency prediction for estimation of a single parameter - $T_1$ using DESPOT1 (Christensen *et al.*, 1974). One healthy volunteer (male, age 25) was scanned on 3T Achieva MRI systems (Best, Netherland) using a 32-channel head coil. Brain images at a resolution $1 \times 1 \times 3mm$ were obtained using 3D Cartesian encoding of a transverse slab with 7 slices, such that the middle slice could be analyzed free of slice profile effects; no parallel imaging acceleration was used. The acquisition consisted of 10 repeats of 6 SPGR sequences with $\alpha = \{5°, 8°, 10°, 13°, 15°, 18°\}$ and $TR = 20ms$, yielding a maximum root-mean-squared RF field of $0.46\mu T$ for $\alpha = 18°$ to minimize bias from Magnetization Transfer (MT) effects (Ou and Gochberg, 2008; Teixeira, Malik and Hajnal, 2019). To compare multiple examples with different efficiencies, all combinations of at least 3 SPGR were considered. Additionally, a transmit field ($B_1^+$) map was acquired using actual flip angle imaging (AFI) method (Yarnykh, 2007) with an isotropic resolution of $5mm$.

All numerical simulations and analyses were performed in a workstation with 64GB of RAM and with an Intel Xeon E5-2687W 0 @ 3.10GHz, using MATLAB R2017b (The MathWorks, Natick, MA, USA) with some functions implemented in C++/MEX using the Eigen linear algebra library (Guënnebaud and Jacob, 2010).

# Results

## Efficiency comparison

Figure 2 shows a comparison of optimized efficiencies for steady-state and transient methods, whilst the optimized acquisition settings are in Supporting Information Table S2 and

Figure S1. Figure 2(a,b) show the distribution of efficiency values over different $T_1$ and $T_2$ values and averaged over off-resonance, while Figure 2(c,d) show spread over different off-resonance values averaged over $T_1$ and $T_2$ values. Consistently we see that the steady-state methods are less efficient than their transient counterparts; DESPOT/JSR is the most efficient steady-state method while balanced MRF starting from thermal equilibrium is the most efficient transient method. In general the best transient method is approximately 3 to 3.5 times more efficient than the best steady-state method. The transient methods have an apparently greater spread in efficiency as a function of $T_1$ and $T_2$. Only the methods that include balanced readouts are sensitive to off-resonance, and of these the transient methods seem more sensitive than the steady-state ones. Nevertheless, methods using balanced readouts are more efficient than those using spoiled readouts despite having to estimate two additional nuisance parameters. Supporting Information Figure S2 compares efficiency of different optimized MRF trains with different numbers of measurements.

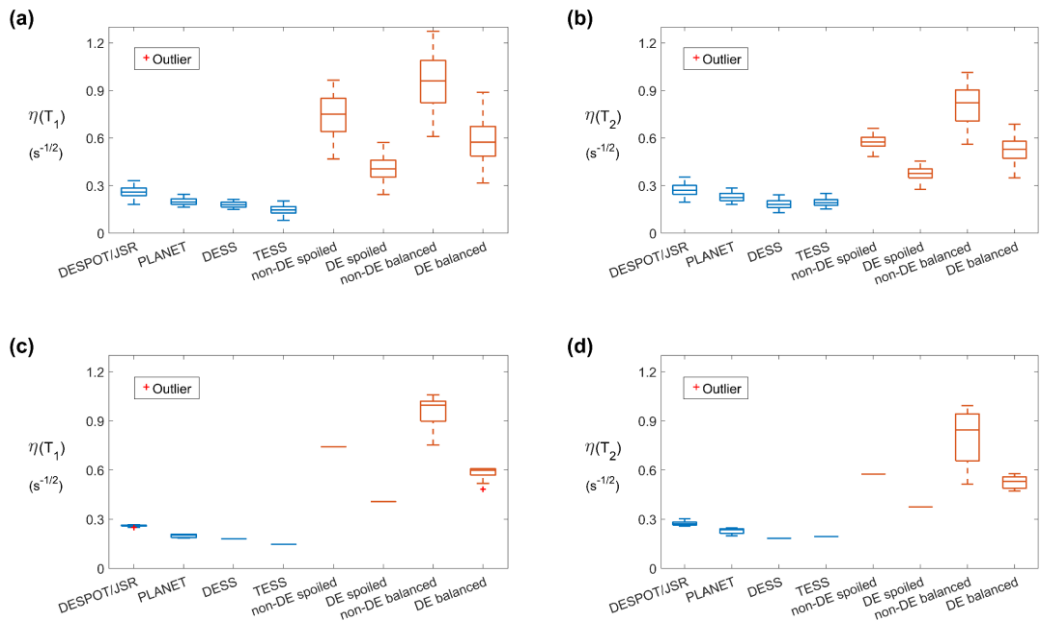

Figure 2: Efficiency comparison for steady-state (blue) and transient (orange) methods as described in subsection 'Efficiency comparison'; in each case the results shown are for the most efficient acquisition of each method (Table 1). (a,b) $T_1, T_2$ efficiency averaged over all off-resonance values; spread corresponds to variability over $\{T_1, T_2\}$. (c,d) $T_1, T_2$ efficiency averaged over all $\{T_1, T_2\}$; spread corresponds to variability over off-resonance frequencies. As may be expected the balanced sequences show greater sensitivity to off-resonance.

Figure 3 shows the results of the Monte Carlo investigation of the dynamics factor $d_R$ for zero-filled reconstructions computed with random and spiral sampling. For this special case, $d_R$ increases quickly for lower undersampling factors R, but then plateaus at higher R;

the level it reaches is directly proportional to $SNR_{image}$ (Figure 3(b)), with spiral sampling achieving lower $d_R$ values than random sampling. Figure 3(c) plots the 'aliasing-to-signal ratio' as a function of R for a scenario with zero thermal noise; empirical fits to data show that to a good approximation this is proportional to $\sqrt{R-1}$ for both sampling schemes used. The full $d_R$ maps are in Supporting Information Figure S3.

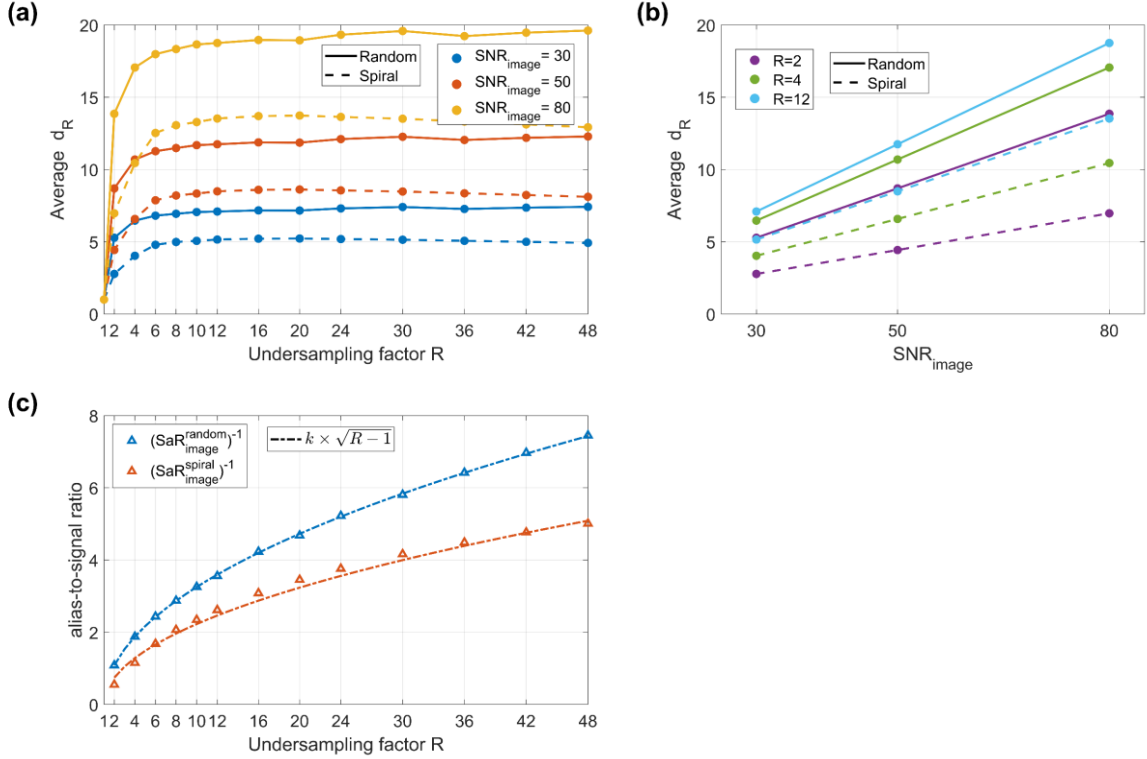

Figure 3: Analysis of the effect of undersampling for random and spiral sampling. (a) Average $d_R$ in the non-zero locations of the Shepp-Logan phantom as a function of the undersampling factor R and (b) as a function of the SNR in the image domain. (c) Aliasing-to-signal ratio as a function of the undersampling factor R and its empirical fit to the expression $k \times \sqrt{R-1}$.

In-vivo validation

Figure 4 depicts comparisons of $T_1$ efficiency for DESPOT1 estimation in-vivo for several subsets of the total dataset. Figure 4(a) shows efficiency maps for a selection of combinations (the full set of maps is in Supporting Information Figure S4). Figure 4(c) shows the average of the experimental $T_1$ efficiency inside the white and gray matter regions (Figure 4(b)) plotted against the average of the theoretical efficiency in the same regions for each combination tested. The correlation coefficient for this comparison is 0.998.

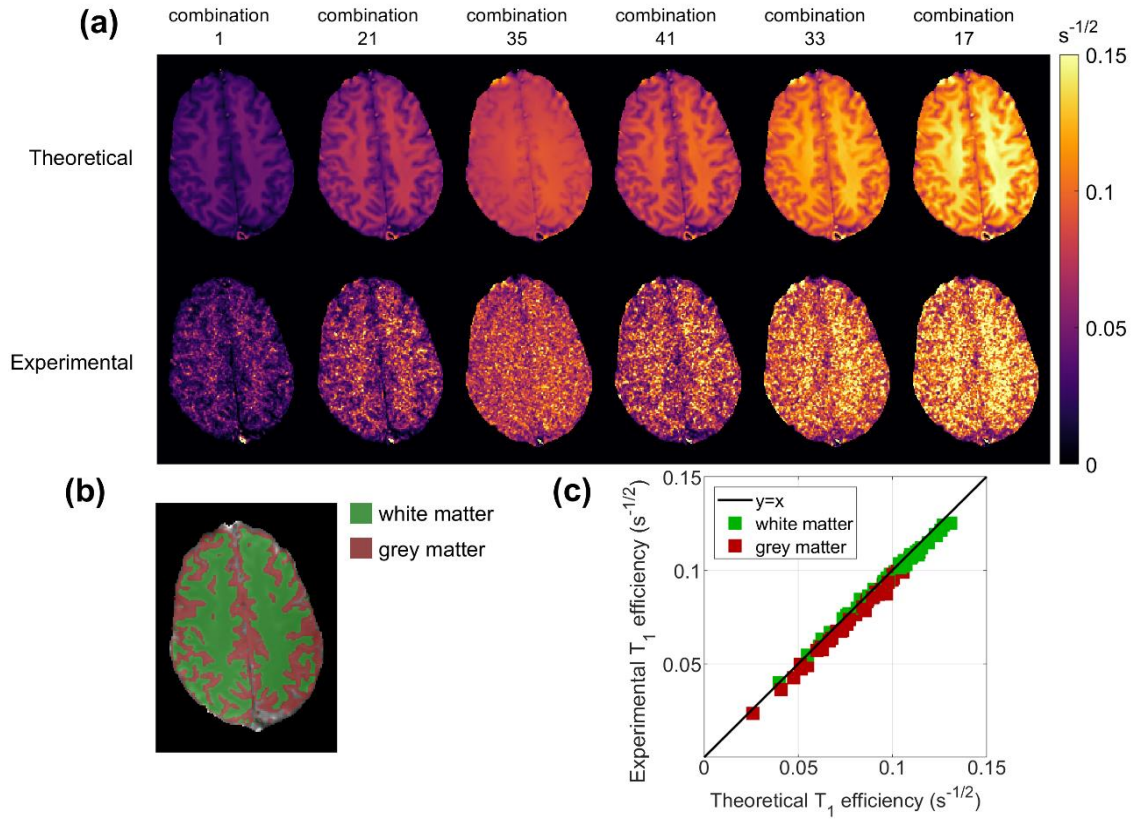

Figure 4: In-vivo validation results for the $T_1$ efficiency of DESPOT1. (a) Theoretical and experimental $T_1$ efficiency maps obtained for some combinations of the acquired SPGRs. (b) White and gray matter masks obtained using FSL FAST (Zhang, Y; Brady, M; Smith, 2001). (c) Average experimental $T_1$ efficiency inside the gray and white matter masks plotted against the respective theoretical efficiency for each combination of SPGRs. A table with all SPGR combinations is provided in Supporting Table S3 and all maps are in Supporting Information Figure S4. The correction for incomplete spoiling in SPGR proposed by Baudrexel et al. (Baudrexel *et al.*, 2018) was implemented. Brain extraction was performed using FSL BET (Smith, 2002) and all images were registered using MIRTK (Schuh, Rueckert and Schnabel, 2013). Gold standard values for $T_1$ and $M_0$ were estimated from fitting to the average of all repetitions and the noise standard deviation $\sigma_0$ was estimated from the image domain (Supporting Information Figure S5).

# Discussion

This work presented a comparison of qMRI methods for simultaneous $T_1$ and $T_2$ estimation based on an efficiency metric, $\eta$, that quantifies the information encoded about the parameters per square root of acquisition time. The efficiency metric was validated in-vivo using the DESPOT1 method (Figure 4). The results showed an excellent correlation between experimental and theoretical $T_1$ efficiency (correlation coefficient 0.998) when averaged over white/gray matter masks.

The metric $\eta$ considers intrinsic efficiency related to the dynamics of the spin system only. Therefore, it considers what mainly distinguishes one method from another: the intrinsic way magnetization is manipulated to achieve greater or lesser sensitivity to the tissue parameters of interest. Nevertheless, according to Eq. [2] there are two other factors that affect the final parameter-to-noise-ratio, and these do need to be considered in the bigger picture. One factor is the amount of data acquired, as efficiency is normalized by $\sqrt{T_{acq}}$. Thus, the SNR advantage yielded by 3D encoding compared to 2D encoding should be accounted for when comparing the expected parameter-to-noise-ratio. The other is the intrinsic SNR of the experiment, specified by $SNR_{max}$. While increasing the repetition time might appear to reduce efficiency, one could for example make use of that time to reduce the receiver bandwidth, thus increasing $SNR_{max}$. Moreover, different *k*-space sampling strategies could be employed to make more efficient use of that time, like spiral sampling (Jiang *et al.*, 2015) or even EPI (Rieger *et al.*, 2017, 2018). However, there are pragmatic limits to which such trade-off can be explored due to $T_2^*$ decay and increased sensitivity to water-fat shift, off-resonance, among others. In the end, intrinsic SNR, amount of data acquired, and efficiency have to weighed together to determine the best method for a specific application.

A Cramér-Rao approach was adopted for determining the uncertainty of parameter estimation, providing a secure mathematical basis for interpretation of results with a clear domain of applicability based on unbiased estimators (Sengupta and Kay, 1995). This connects to a key assumption in relaxometry - the biophysical model. In this work we assumed that the magnetization dynamics are well represented by the Bloch equations and that each voxel contains only one tissue type. In practice this ignores factors such as magnetization transfer (Ou and Gochberg, 2008; Teixeira *et al.*, 2019; Teixeira, Malik and Hajnal, 2019) and diffusion (Kobayashi and Terada, 2018) that can produce systematic differences between the model and data, biasing parameter estimation, and to which different methods have different sensitivities.

Figure 2 suggests that after optimization, the transient methods are generally more efficient than the steady-state methods, at least when there is no undersampling. From the latter, DESPOT/JSR has the highest optimized efficiency. This approach combines SPGR and balanced SSFP sequences, giving it a large degree of flexibility. We found that the most efficient acquisitions consisted of mainly bSSFP sequences but having at least one SPGR enhances efficiency by decorrelating $T_1$ and $T_2$ information. The PLANET method uses only bSSFP sequences, which results in slightly lower efficiency as it constitutes a constrained case of JSR that excludes SPGR measurements and forces a single flip angle and TR. Optimizing bSSFP sequences to maximize $\eta$ leads to use of multiple different flip angles. Because TESS and DESS obtain multiple echoes per TR period, they are often thought of as efficient, however our results indicate this may not be the case. Although they measure multiple echo pathways, the information in these echoes is correlated since they share the same flip angle and TR, and the higher order echoes often have lower signal amplitudes; it could be more efficient to obtain more diverse data instead.

The transient (MRF) sequences are divided into those in driven equilibrium (DE) and those that start from thermal equilibrium, and between either gradient spoiled or balanced readouts. The results in Figure 2 reported only the most efficient transient acquisition with at least 400 measurements. Supporting Information Figure S2 expands this with efficiencies for different numbers of measurements. It was seen that very high efficiencies can be achieved for short MRF pulse trains that start from thermal equilibrium; these may not be viable for performing the spatial encoding required for 3D but could potentially be used for 2D encoding. Methods that start from thermal equilibrium (i.e. the 'non-DE' sequences) are 'privileged' by assuming sufficient time has elapsed following previous acquisitions to allow full recovery. If this recovery time is also included in the efficiency calculation these methods lose their high efficiency. In the case of longer acquisitions, both MRF sequences in DE and starting from thermal equilibrium converged to a similar efficiency, which is still higher than steady-state methods.

When considering spin dynamics alone, the transient ('MRF like') methods are undoubtedly more efficient. However, this excludes the fact that these methods usually employ a significant degree of undersampling, whose effect can be included by computing dynamics-factor $d_R$. This factor is dependent on the object, spatial sampling and image reconstruction method used. In order to estimate an approximate value for $d_R$ we focused on zero-filled reconstructions that were initially proposed for MRF (Ma *et al.*, 2013). Figure 3 shows that for higher undersampling factors R, $d_R$ is independent of R but scales linearly with the SNR of the data. To understand these relationships, Figure 3(c) shows empirically that the 'alias-to-signal' ratio scales as $\sqrt{R-1}$. Substituting this into Eq.[8] we find:

$$d_R \approx \sqrt{\frac{k(R-1)}{R}SNR_{image}^2 + 1},$$

where $k$ is a scaling constant, which yields $d_R \propto SNR_{image}$ for large R and large SNR. Though at first unintuitive this should be expected since high SNR implies that the dominant source of 'noise' in the reconstruction comes from aliasing. It is important to note that this analysis looks at the simplest MRF reconstruction (zero filling); current reconstructions (Pierre *et al.*, 2016; Zhao *et al.*, 2016, 2018; Assländer, Cloos, *et al.*, 2018) should be expected to reduce $d_R$ since they attempt to resolve the aliased signal rather than treat it as noise (Stolk and Sbrizzi, 2019) and may employ some prior knowledge or form of regularization (Pierre *et al.*, 2016; Assländer, Cloos, *et al.*, 2018; Zhao *et al.*, 2018). Hence the $d_R$ values computed by our analysis might be considered as upper bounds. However, more complex and/or regularized reconstructions may introduce bias, which is beyond the scope of the proposed efficiency metric.

Our comparison only considered gradient-echo based methods that simultaneously estimate $T_1$ and $T_2$ but could be used for other methods and other parameters. We have considered precision as the only evaluation criterion but did not account for any other reason why one method might be preferred to another – for example (in)sensitivity to $B_1^+$ inhomogeneities, lower SAR or motion sensitivity. These considerations can be incorporated by changing the model and/or constraining the design space, which might change the comparison landscape obtained here.

# Conclusion

This work presents a comparison of several qMRI methods based on an efficiency metric that is the ratio of the best-case parameter-to-noise ratio to the maximum achievable SNR, normalized to the square root of the acquisition time. This allows the performance of different qMRI methods to be quantified and optimized, and has been used here to compare a range of well-established methods that simultaneously estimate $T_1$ and $T_2$. We found that transient qMRI sequences have the potential to be 3 to 3.5 times more efficient than steady-state alternatives for both $T_1$ and $T_2$ mapping. Furthermore, methods based on balanced readouts outperformed methods based on spoiled readouts. The impact of undersampling on the efficiency was analyzed for the sub-optimal zero-filled reconstruction that treats aliasing artifacts as noise, and a drop in efficiency by a factor of 5 could be easily attained in practice, meaning that current more advanced reconstructions are important to realize the gains offered by MRF.

# Acknowledgements

This work was supported by the Wellcome/EPSRC Centre for Medical Engineering (WT 203148/Z/16/Z) and by the National Institute for Health Research (NIHR) Biomedical Research Centre based at Guy's and St Thomas' NHS Foundation Trust and King's College London and/or the NIHR Clinical Research Facility, and funded by the King's College London & Imperial College London EPSRC Centre for Doctoral Training in Medical Imaging (EP/L015226/1). The views expressed are those of the author(s) and not necessarily those of the NHS, the NIHR or the Department of Health and Social Care.

# Data availability

The code that supports the findings of this study is openly available at the following URL: https://github.com/mriphysics/qMRI_efficiency (hash f311785 was the version at time of submission). The in-vivo data generated and/or analysed during the current study are not publicly available for legal/ethical reasons.

# Ethical statement

Healthy volunteer gave written informed consent according to local ethics requirements (Guy's Research Ethics Committee, ID 01/11/12).

# References

Assländer, J., Novikov, D. S., *et al.* (2018) 'Hybrid-State Free Precession in Nuclear Magnetic Resonance'. Available at: https://doi.org/10.1038/s42005-019-0174-0 (Accessed: 14 August 2019).

Assländer, J., Cloos, M. A., *et al.* (2018) 'Low rank alternating direction method of multipliers reconstruction for MR fingerprinting', *Magnetic Resonance in Medicine*, 79(1), pp. 83–96. doi: 10.1002/mrm.26639.

Assländer, J. *et al.* (2019) 'Optimized quantification of spin relaxation times in the hybrid state', *Magnetic Resonance in Medicine*, (April), pp. 1385–1397. doi: 10.1002/mrm.27819.

Assländer, J. (2020) 'A perspective on MR fingerprinting', *Journal of Magnetic Resonance Imaging*, pp. 1–10. doi: 10.1002/jmri.27134.

Assländer, J., Glaser, S. J. and Hennig, J. (2017) 'Pseudo Steady-State Free Precession for MR-Fingerprinting', *Magnetic Resonance in Medicine*, 77(3), pp. 1151–1161. doi: 10.1002/mrm.26202.

Baudrexel, S. *et al.* (2018) 'T 1 mapping with the variable flip angle technique: A simple correction for insufficient spoiling of transverse magnetization', *Magnetic Resonance in Medicine*, 79(6), pp. 3082–3092. doi: 10.1002/mrm.26979.

Bojorquez, J. Z. *et al.* (2017) 'What are normal relaxation times of tissues at 3 T?', *Magnetic Resonance Imaging*, 35, pp. 69–80. doi: 10.1016/j.mri.2016.08.021.

Christensen, K. A. *et al.* (1974) 'Optimal determination of relaxation times of Fourier transform nuclear magnetic resonance. Determination of spin-lattice relaxation times in chemically polarized species', *Journal of Physical Chemistry*, 78(19), pp. 1971–1976. doi: 10.1021/j100612a022.

Crawley, A. P. and Henkelman, R. M. (1988) 'A comparison of one-shot and recovery methods in T1 imaging', *Magnetic Resonance in Medicine*, 7(1), pp. 23–34. doi: 10.1002/mrm.1910070104.

Deoni, S. C. L., Rutt, B. K. and Peters, T. M. (2003) 'Rapid combined T1 and T2 mapping using gradient recalled acquisition in the steady state', *Magnetic Resonance in Medicine*, 49(3), pp. 515–526. doi: 10.1002/mrm.10407.

Edelstein, W. A. *et al.* (1983) 'Signal, noise, and contrast in nuclear magnetic resonance (NMR) imaging', *Journal of Computer Assisted Tomography*, pp. 391–401. doi: 10.1097/00004728-198306000-00001.

Glover, G. H. (1999) 'Simple analytic spiral K-space algorithm', *Magnetic Resonance in Medicine*, 42(2), pp. 412–415. doi: 10.1002/(SICI)1522-2594(199908)42:2<412::AID-MRM25>3.0.CO;2-U.

Gras, V. *et al.* (2017) 'Diffusion-weighted DESS protocol optimization for simultaneous mapping of the mean diffusivity, proton density and relaxation times at 3 Tesla', *Magnetic Resonance in Medicine*, 78(1), pp. 130–141. doi: 10.1002/mrm.26353.

Guënnebaud, G. and Jacob, B. (2010) 'Eigen v3'. Available at: http://eigen.tuxfamily.org.

Heule, R., Ganter, C. and Bieri, O. (2014) 'Triple echo steady-state (TESS) relaxometry', *Magnetic Resonance in Medicine*, 71(1), pp. 230–237. doi: 10.1002/mrm.24659.

Hu, C. and Peters, D. C. (2019) 'SUPER: A blockwise curve-fitting method for accelerating MR parametric mapping with fast reconstruction', *Magnetic Resonance in Medicine*, 81(6), pp. 3515–3529. doi: 10.1002/mrm.27662.

Jiang, Y. *et al.* (2015) 'MR fingerprinting using fast imaging with steady state precession (FISP) with spiral readout', *Magnetic Resonance in Medicine*, 74(6), pp. 1621–1631. doi: 10.1002/mrm.25559.

Jones, J. A. *et al.* (1996) 'Optimal sampling strategies for the measurement of spin-spin relaxation times', *Journal of Magnetic Resonance - Series B*, 113(1), pp. 25–34. doi: 10.1006/jmrb.1996.0151.

Kobayashi, Y. and Terada, Y. (2018) 'Diffusion-weighting Caused by Spoiler Gradients in the Fast Imaging with Steady-state Precession Sequence May Lead to Inaccurate T2 Measurements in MR Fingerprinting', *Magnetic Resonance in Medical Sciences*, pp. 1–9. doi: 10.2463/mrms.tn.2018-0027.

Ma, D. *et al.* (2013) 'Magnetic resonance fingerprinting'. doi: 10.1038/nature11971.

Ma, D. *et al.* (2018) 'Fast 3D magnetic resonance fingerprinting for a whole-brain coverage', *Magnetic Resonance in Medicine*, 79(4), pp. 2190–2197. doi: 10.1002/mrm.26886.

Nataraj, G., Nielsen, J. and Fessler, J. A. (2017) 'Optimizing MR Scan Design for Model-Based', 36(2), pp. 467–477.

O'Donnell, M., Gore, J. C. and Adams, W. J. (1986) 'Toward an automated analysis system for nuclear magnetic resonance imaging. I. Efficient pulse sequences for simultaneous T1–T2 imaging', *Medical Physics*, 13(2), pp. 182–190. doi: 10.1118/1.595943.

Ou, X. and Gochberg, D. F. (2008) 'MT effects and T1 quantification in single-slice spoiled gradient echo imaging', *Magnetic Resonance in Medicine*, 59(4), pp. 835–845. doi: 10.1002/mrm.21550.

Pierre, E. Y. *et al.* (2016) 'Multiscale reconstruction for MR fingerprinting', *Magnetic Resonance in Medicine*, 75(6), pp. 2481–2492. doi: 10.1002/mrm.25776.

Pipe, J. G. and Zwart, N. R. (2014) 'Spiral trajectory design: A flexible numerical algorithm and base analytical equations', *Magnetic Resonance in Medicine*, 71(1), pp. 278–285. doi: 10.1002/mrm.24675.

Pruessmann, K. P. *et al.* (1999) 'SENSE: Sensitivity encoding for fast MRI', *Magnetic Resonance in Medicine*, 42(5), pp. 952–962. doi: 10.1002/(SICI)1522-2594(199911)42:5<952::AID-MRM16>3.0.CO;2-S.

Rieger, B. *et al.* (2017) 'Magnetic resonance fingerprinting using echo-planar imaging: Joint quantification of T1 and T2∗ relaxation times', *Magnetic Resonance in Medicine*, 78(5), pp. 1724–1733. doi: 10.1002/mrm.26561.

Rieger, B. *et al.* (2018) 'Time efficient whole-brain coverage with MR Fingerprinting using slice-interleaved echo-planar-imaging', *Scientific Reports*, 8(1), pp. 1–12. doi: 10.1038/s41598-018-24920-z.

Roemer, P. B. *et al.* (1990) 'The NMR phased array', *Magnetic Resonance in Medicine*, 16(2), pp. 192–225. doi: 10.1002/mrm.1910160203.

Sbrizzi, A. *et al.* (2017) *Dictionary-free MR Fingerprinting reconstruction of balanced-GRE sequences*. Available at: https://arxiv.org/pdf/1711.08905.pdf (Accessed: 14 August 2019).

Schuh, A., Rueckert, D. and Schnabel, J. (2013) *Medical Image Registration ToolKit (MIRTK)*. Available at: https://mirtk.github.io/.

Sengupta, S. K. and Kay, S. M. (1995) 'Fundamentals of Statistical Signal Processing: Estimation Theory', *Technometrics*, 37(4), p. 465. doi: 10.2307/1269750.

Shcherbakova, Y. *et al.* (2018) 'PLANET: An ellipse fitting approach for simultaneous T 1 and T 2 mapping using phase-cycled balanced steady-state free precession', *Magnetic Resonance in Medicine*, 79(2), pp. 711–722. doi: 10.1002/mrm.26717.

Smith, S. M. (2002) 'Fast robust automated brain extraction', *Human Brain Mapping*, 17(3), pp. 143–155. doi: 10.1002/hbm.10062.

Stolk, C. C. and Sbrizzi, A. (2019) 'Understanding the combined effect of k-space undersampling and transient states excitation in MR Fingerprinting reconstructions', *IEEE Transactions on Medical Imaging*, pp. 1–1. doi: 10.1109/tmi.2019.2900585.

Teixeira, R. P. A. G. *et al.* (2019) 'Controlled saturation magnetization transfer for reproducible multivendor variable flip angle T1 and T2 mapping', *Magnetic Resonance in Medicine*, (October), pp. 1–16. doi: 10.1002/mrm.28109.

Teixeira, R. P. A. G., Malik, S. J. and Hajnal, J. V. (2017) 'Joint system relaxometry (JSR) and Cramer-Rao lower bound optimization of sequence parameters: A framework for enhanced precision of DESPOT T1 and T2 estimation', *Magnetic Resonance in Medicine*. doi: 10.1002/mrm.26670.

Teixeira, R. P. A. G., Malik, S. J. and Hajnal, J. V. (2019) 'Fast quantitative MRI using controlled saturation magnetization transfer', *Magnetic Resonance in Medicine*, 81(2), pp. 907–920. doi: 10.1002/mrm.27442.

van Valenberg, W. *et al.* (2017) 'Determining the Time Efficiency of Quantitative MRI Methods using Bloch Simulations', in, p. 1470. Available at: http://indexsmart.mirasmart.com/ISMRM2017/PDFfiles/1470.html (Accessed: 29 September 2019).

Welsch, G. H. *et al.* (2009) 'Rapid estimation of cartilage T2 based on double echo at steady state (DESS) with 3 Tesla', *Magnetic Resonance in Medicine*, 62(2), pp. 544–549. doi: 10.1002/mrm.22036.

Wundrak, S. *et al.* (2016) 'Golden Ratio Sparse MRI Using Tiny Golden Angles', *Magnetic Resonance in Medicine*, 2378(March 2015), pp. 2372–2378. doi: 10.1002/mrm.25831.

Yarnykh, V. L. (2007) 'Actual flip-angle imaging in the pulsed steady state: A method for rapid three-dimensional mapping of the transmitted radiofrequency field', *Magnetic Resonance in Medicine*, 57(1), pp. 192–200. doi: 10.1002/mrm.21120.

Zhang, Y; Brady, M; Smith, S. (2001) 'Segmentation of brain MR images through a hidden Markov random field model and the expectation-maximization algorithm', *IEEE transactions on medical imaging*, 20(1), pp. 45–57.

Zhao, B. *et al.* (2016) 'Maximum Likelihood Reconstruction for Magnetic Resonance Fingerprinting', *IEEE TRANSACTIONS ON MEDICAL IMAGING*, 35(8), pp. 1812–1823. doi: 10.1109/TMI.2016.2531640.

Zhao, B. *et al.* (2018) 'Improved magnetic resonance fingerprinting reconstruction with low-rank and subspace modeling', *Magnetic Resonance in Medicine*, 79(2), pp. 933–942. doi: 10.1002/mrm.26701.

Zhao, B. *et al.* (2019) 'Optimal experiment design for magnetic resonance fingerprinting: Cramér-rao bound meets spin dynamics', *IEEE Transactions on Medical Imaging*, 38(3), pp. 844–861. doi: 10.1109/TMI.2018.2873704.

# Supporting Figures and Tables

**Table of contents:**

**Supporting Information Table S1**

| Method | Signal model | Decision variables | Lower and upper bounds | Non-linear constraints |
|---|---|---|---|---|
| DESPOT/JSR | $M_{xy}^{SPGR} = \sin(\alpha)\, e^{-TE/T_2}\, \dfrac{1-e^{-\frac{TR}{T_1}}}{1-\cos(\alpha)\,e^{-\frac{TR}{T_1}}}\, M_0 e^{i(\phi_0+2\pi\omega_0 TE)}$ <br><br> $M_{xy}^{bSSFP} = \dfrac{M_0 e^{i(\phi_0+\pi\omega_0 TR)} \sin(\alpha)\, e^{-\frac{TR}{2T_2}}\left(1-e^{-\frac{TR}{T_1}}\right)\left(1-e^{-\frac{TR}{T_2}} e^{i(2\pi\omega_0 TR+\Phi_{inc})}\right)}{1 - e^{-\frac{TR}{T_1}}\cos(\alpha) - e^{-\frac{2TR}{T_2}}\left(e^{-\frac{TR}{T_1}} - \cos(\alpha)\right) - d\cdot\cos(2\pi\omega_0 TR+\Phi_0)}$ <br><br> $d = e^{-\frac{TR}{T_2}}\left(1-e^{-\frac{TR}{T_1}}\right)(1+\cos(\alpha))$ | $\alpha_i, TR_i, TE_i,$ <br> $i=1,\dots,N_{SPGR}$ <br> $\alpha_i, TR_i, \Phi_{inc,i},$ <br> $i=1,\dots,N_{bSSFP}$ | $0° \leq \alpha_i \leq 90°$ <br> $5ms \leq TR_i$ <br> $2ms \leq TE_i \leq TR_i - 2ms$ <br> $0° \leq \Phi_{inc,i} \leq 360°$ | - |
| PLANET | Same equation as in DESPOT/JSR for $M_{xy}^{bSSFP}$ | $\alpha, TR, \Phi_{inc,i},$ <br> $i=1,\dots,N_{bSSFP}$ | $0° \leq \alpha \leq 90°$ <br> $5ms \leq TR$ <br> $0° \leq \Phi_{inc,i} \leq 360°$ | - |
| DESS | EPG simulation for single pool model (no diffusion); uses $\tilde{F}_-(0)$ and $\tilde{F}_-(1)$ states ($TE_1$ and $TE_2$, respectively) after reaching steady-state | $\alpha_i, TR_i,$ <br> $TE_{1,i}, TE_{2,i},$ <br> $i=1,\dots,N_{DESS}$ | $0° \leq \alpha_i \leq 90°$ <br> $10ms \leq TR_i$ <br> $2ms \leq TE_{1,i} \leq TE_{2,i} - 2ms$ <br> $TE_{1,i} + 2ms \leq TE_{2,i} \leq TR_i - 2ms$ | - |
| TESS | EPG simulation for single pool model (no diffusion); uses $\tilde{F}_-(0)$, $\tilde{F}_-(1)$ and $\tilde{F}_+(1)$ states ($TE_1$, $TE_2$ and $TE_3$, respectively) after reaching steady-state | $\alpha_i, TR_i,$ <br> $TE_{1,i}, TE_{2,i}, TE_{3,i},$ <br> $i=1,\dots,N_{TESS}$ | $0° \leq \alpha_i \leq 90°$ <br> $15ms \leq TR_i$ <br> $2ms \leq TE_{1,i} \leq TE_{2,i} - 2ms$ <br> $TE_{1,i} + 2ms \leq TE_{2,i} \leq TE_{3,i} - 2ms$ <br> $TE_{2,i} + 2ms \leq TE_{3,i} \leq TR_i - 2ms$ | - |
| (non-DE and DE) spoiled MRF | EPG simulation for single pool model (no diffusion). For DE implementation, the same sequence is repeated in a loop for at least 3 times or $10 \times T_1$ to ensure DE is reached and then the signals from the last cycle are used | $\alpha_i, TR_i,$ <br> $i=1,\dots,N$ | $0° \leq \alpha_1 \leq 180°$ <br> $0° \leq \alpha_i \leq 90°,$ <br> $i=2,\dots,N$ <br> $5ms \leq TR_i$ | - |
| (non-DE and DE) balanced MRF | Isochromat simulation for single pool model (no diffusion). For DE implementation, the same sequence is repeated in a loop for at least 3 times or $10 \times T_1$ to ensure DE is reached and then the signals from the last cycle are used; furthermore, perfect spoiling is assumed in between cycles | $\vartheta_i,$ <br> $i=1,\dots,N$ | $\vartheta_1 = 0°$ <br> $0° \leq \vartheta_i \leq 45°,$ <br> $i=2,\dots,N$ <br> $\vartheta_i = \alpha_i, \quad i=1$ <br> $\vartheta_i = \alpha_i + \alpha_{i-1},\ i \geq 2$ | $\|\vartheta_{i+1} - \vartheta_i\| \leq 5°, i \geq 1$ <br><br> $\|\vartheta_{i+1} - 2\vartheta_i + \vartheta_{i-1}\| \leq 0.5°, i \geq 2$ |

**Supporting Information Table S1:** Signal models, acquisition settings (decision variables) optimized and their respective lower/upper bounds and non-linear constraints for each method analyzed. For PLANET the ellipse description proposed for data fitting by Shcherbakova et al. (Shcherbakova et al., 2018) was not considered here. Instead, the steady-state equation for bSSFP was used. Balanced MRF was described using the polar angle $\vartheta$ description as introduced by Assländer et al. (Assländer, Glaser and Hennig, 2017) and the repetition time was fixed at $TR = 5ms$. Combined with a forced smooth flip angle variation, this aims to ensure that a spin echo will be formed every $TE$ (Assländer, Glaser and Hennig, 2017); the values on the non-linear constraints imposed onto $\vartheta$ were heuristically found to provide a smooth $\vartheta$ variation. Furthermore, to minimize signal oscillations for big off-resonance values the flip angle is forced to start from zero (Sbrizzi et al., 2017), and at the beginning of the sequence the magnetization was assumed to be inverted ($M(t_i) = -M_0$ for non-DE scenario, $M(t_i) = -M(t_f)$ for DE scenario). Glossary: $\alpha$ is the flip angle, $TR$ is the repetition time, $TE$ is the echo time, $\Phi_{inc}$ is the RF phase increment for balanced sequences and $N_*$ is the number of measurements.

**Supporting Information Table S2**: Optimized settings for each of the steady-state methods used for comparison. In each case multiple different configurations were tried – for example DESPOT/JSR can use combinations of SPGR and bSSFP images and many different combinations were tested. Reported here are the best performing (most efficient) acquisition parameters for each method, with details of listed in the table. Full source code can be found online at https://github.com/mriphysics/qMRI_efficiency for replication of these experiments; all details can be found within that code.

| Method | Important details | Settings |
|---|---|---|
| DESPOT/JSR | Multiple versions using different numbers of SPGR and bSSFP acquisitions were trialled. Most efficient used 1 SPGR and 7 bSSFP images | $FA^{SPGR_1} = 35°, TR^{SPGR_1} = 42ms, TE^{SPGR_1} = 2ms$<br>$FA^{bSSFP_1} = 11°, TR^{bSSFP_1} = 5ms, \Phi_{inc}^{bSSFP_1} = 114°$<br>$FA^{bSSFP_2} = 12°, TR^{bSSFP_2} = 5ms, \Phi_{inc}^{bSSFP_2} = 353°$<br>$FA^{bSSFP_3} = 12°, TR^{bSSFP_3} = 5ms, \Phi_{inc}^{bSSFP_3} = 206°$<br>$FA^{bSSFP_4} = 50°, TR^{bSSFP_4} = 5ms, \Phi_{inc}^{bSSFP_4} = 308°$<br>$FA^{bSSFP_5} = 12°, TR^{bSSFP_5} = 5ms, \Phi_{inc}^{bSSFP_5} = 161°$<br>$FA^{bSSFP_6} = 13°, TR^{bSSFP_6} = 5ms, \Phi_{inc}^{bSSFP_6} = 320°$<br>$FA^{bSSFP_7} = 49°, TR^{bSSFP_7} = 5ms, \Phi_{inc}^{bSSFP_7} = 192°$ |
| PLANET | Multiple versions using different numbers of bSSFP acquisitions were trialled. Most efficient used 20 bSSFP images | $FA^{bSSFP} = 6°, TR^{bSSFP} = 7ms$<br>$\Phi_{inc}^{bSSFP_i} = \{0°, 20°, 41°, 61°, 81°, 102°, 118°, 133°,$<br>$149°, 165°, 180°, 195°, 211°, 227°, 242°, 258°,$<br>$279°, 299°, 319°, 340°\}$ |
| DESS | Multiple versions using different numbers of gradient spoiled sequences without RF spoiling were trialled. Most efficient used 2 DESS acquisitions | $FA^{DESS_1} = 12°, TR^{DESS_1} = 11ms$<br>$TE_1^{DESS_1} = 2ms, TE_2^{DESS_1} = 5ms$<br>$FA^{DESS_2} = 39°, TR^{DESS_2} = 25ms$<br>$TE_1^{DESS_2} = 2ms, TE_2^{DESS_2} = 21ms$ |
| TESS | Multiple versions using different numbers of gradient spoiled sequences without RF spoiling were trialled. Most efficient used 1 TESS acquisition | $FA^{TESS_1} = 20°, TR^{TESS_1} = 15ms$<br>$TE_1^{TESS_1} = 2ms, TE_2^{TESS_1} = 4ms, TE_3^{TESS_1} = 13ms$ |

**Supporting Information Figure S1**: Optimized settings for each of the transient methods used for comparison. For each method multiple acquisitions with different flip angle train lengths were individually optimized and here are the best performing (most efficient) acquisition parameters for each method with at least 400 measurements. Optimal flip angle train and TRs for (a) non-DE spoiled MRF and (b) DE spoiled MRF. Optimal flip angle train for (c) non-DE balanced MRF and (d) DE balanced MRF. Full source code can be found online at https://github.com/mriphysics/qMRI_efficiency for replication of these experiments; all details can be found within that code.

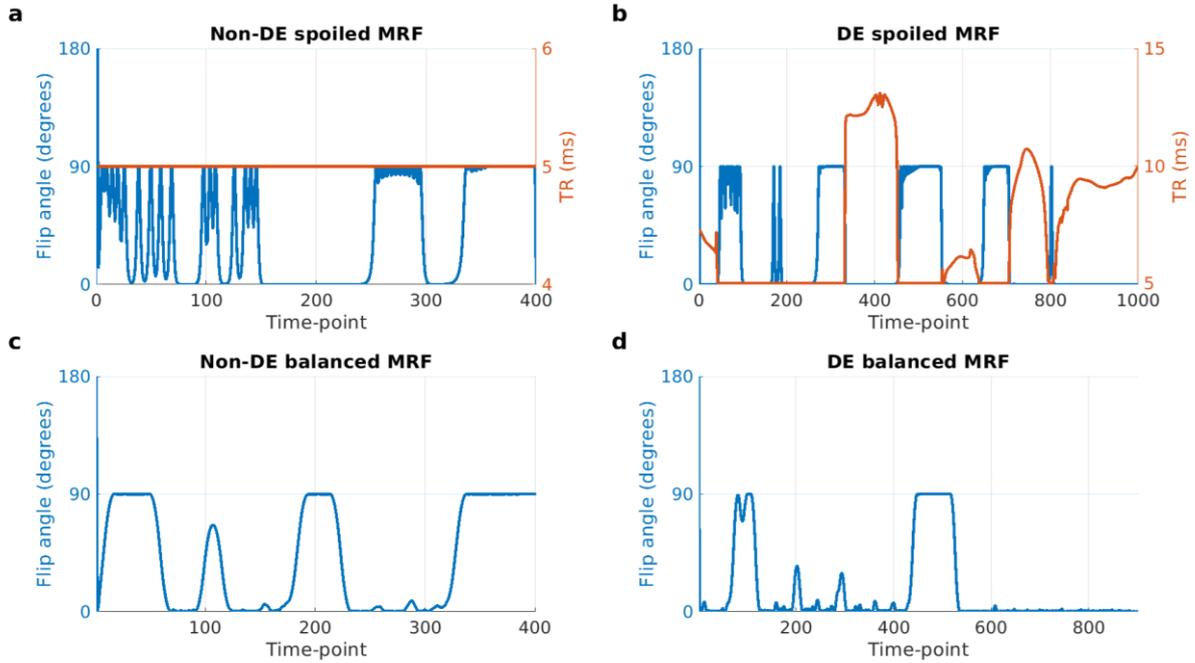

**Supporting Information Figure S2**: Efficiency of each individually optimized fingerprint (Table 1) plotted against its acquisition time $T_{acq}$ for cases where the magnetization is initially in thermal equilibrium. (a,c) and where a recovery time of $T_{recovery} = 5s\ (\approx 5 \times T_1)$ is added to the non-Driven Equilibrium (DE) MRF methods (b,d). Note that methods starting from thermal equilibrium (a,c) have high efficiency for acquisitions with short $T_{acq}$, peaking at $T_{acq} \approx T_1$ like Assländer (Assländer, 2020) reported for balanced MRF. The current comparison includes spoiled MRF, which is shown to be below the efficiency curve of balanced MRF. The pale markers in (a,c) have less than 400 measurements and were not considered for further analyses due to possible incompatibility with spatial encoding. On the other hand, the markers delineated in black are the most $T_1$ and $T_2$ efficient acquisition considered fur further analysis. The efficiency peak visible for non-DE methods (a,c) is lost if a recovery time is added (b,d) to ensure they return to thermal equilibrium. In the limit of long $T_{acq}$, where the different initial conditions weigh less, both DE and non-DE approaches in spoiled and balanced MRF converge to the same efficiency, as expected. This asymptotic efficiency is still superior to that of steady-state methods.

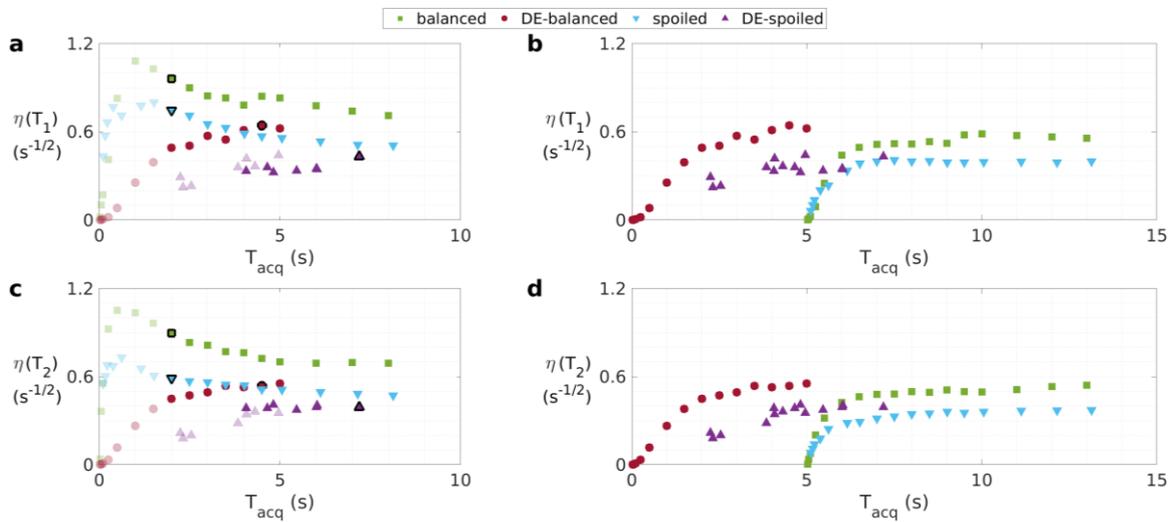

**Supporting Information Figure S3**: Dynamics factor $d_R$ maps obtained for (a) random and (b) spiral sampling, testing different undersampling factors $R$ and SNR levels. For spiral sampling, a uniform radial density Archimedean spiral (Glover, 1999; Pipe and Zwart, 2014) rotated each trial by the tiny golden angle (Wundrak et al., 2016) was used with radial undersampling factor $R$, while for random sampling each k-space location was sampled with probability $1/R$. Note that while for random sampling the dynamics factor maps are uniform, revealing an homogeneous noise amplification, for spiral sampling there is some structure in its maps. This is because the uniform radial density in the spiral causes aliasing fold-over consistently in the same positions. This shows that spiral sampling aliasing may not always be considered to be 'noise-like', which could be alleviated by variable-density spiral.

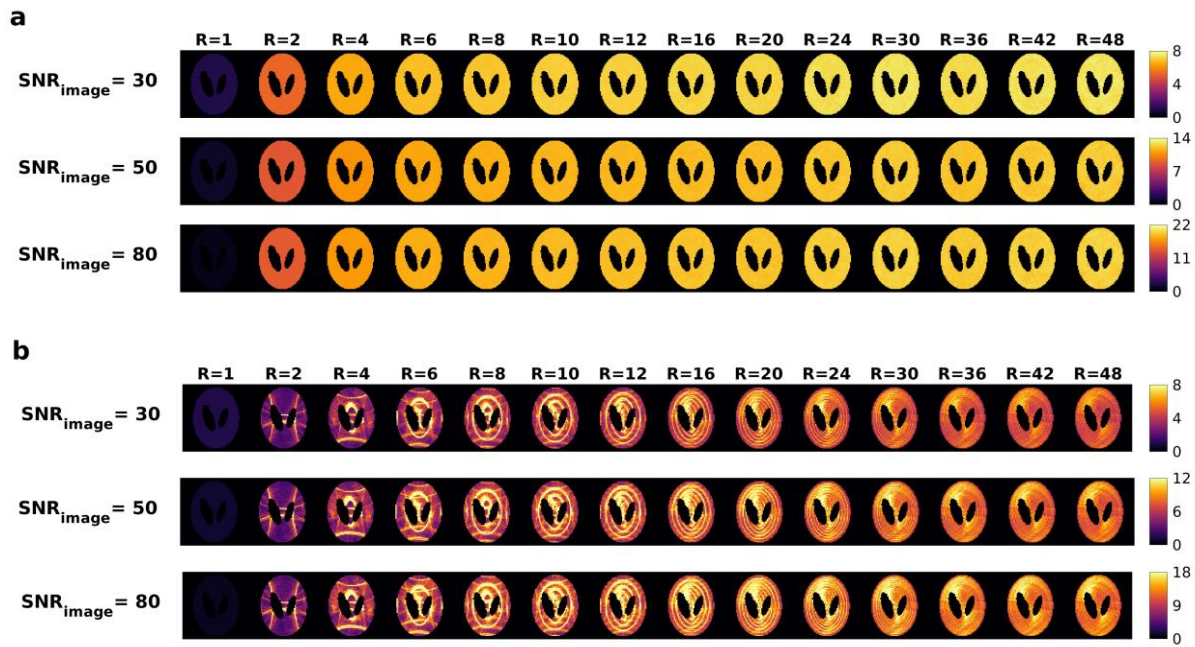

**Supporting Information Table S3**: Numbered flip angle combinations from the acquired DESPOT1 data used to estimate $T_1$ with a range of different estimation efficiencies. All acquisitions used the same TR and other settings, only flip angle differed. Acquisition settings were handcrafted to generate a wide range of efficiencies.

| Combination | Flip angles (deg) | Combination | Flip angles (deg) | Combination | Flip angles (deg) |
|---|---|---|---|---|---|
| 1 | 13, 15, 18 | 15 | 5, 10, 15 | 29 | 5, 10, 13, 15 |
| 2 | 10, 15, 18 | 16 | 5, 10, 13 | 30 | 5, 8, 15, 18 |
| 3 | 10, 13, 18 | 17 | 5, 8, 18 | 31 | 5, 8, 13, 18 |
| 4 | 10, 13, 15 | 18 | 5, 8, 15 | 32 | 5, 8, 13, 15 |
| 5 | 8, 15, 18 | 19 | 5, 8, 13 | 33 | 5, 8, 10, 18 |
| 6 | 8, 13, 18 | 20 | 5, 8, 10 | 34 | 5, 8, 10, 15 |
| 7 | 8, 13, 15 | 21 | 10, 13, 15, 18 | 35 | 5, 8, 10, 13 |
| 8 | 8, 10, 18 | 22 | 8, 13, 15, 18 | 36 | 5, 8, 10, 13, 15 |
| 9 | 8, 10, 15 | 23 | 8, 10, 15, 18 | 37 | 5, 8, 10, 13, 18 |
| 10 | 8, 10, 13 | 24 | 8, 10, 13, 18 | 38 | 5, 8, 10, 15, 18 |
| 11 | 5, 15, 18 | 25 | 8, 10, 13, 15 | 39 | 5, 8, 13, 15, 18 |
| 12 | 5, 13, 18 | 26 | 5, 13, 15, 18 | 40 | 5, 10, 13, 15, 18 |
| 13 | 5, 13, 15 | 27 | 5, 10, 15, 18 | 41 | 8, 10, 13, 15, 18 |
| 14 | 5, 10, 18 | 28 | 5, 10, 13, 18 | 42 | 5, 8, 10, 13, 15, 18 |

**Supporting Information Figure S4**: Experimentally measured and theoretically predicted $T_1$ estimation efficiency for white matter (a) and gray matter (b) for all different parameter combinations listed in Supporting Information Table S2. A subset of these is presented in the paper, Figure 4.

**(a)**

**(b)**

**Supporting Information Figure S5:** (a,b) 'Gold standard' $T_1$ and $M_0$ values were estimated voxel-wise using all flip angles and averaging their repetitions. (c) The maximum SNR in each voxel was estimated by dividing $M_0$ in each voxel by the noise standard deviation $\sigma_0$ estimated for that voxel. As defined in Eq.[3] $\sigma_0$ is the noise standard deviation from one measurement – i.e. the characteristic noise from the receiver when acquiring k-space data, scaled to account for any scaling applied by inverse Fourier transformation. For a fully-sampled 3D scan, as this experiment, this is equivalent to the image domain noise standard deviation divided by the square root of the number of k-space lines. The reason why $SNR_{max}$ appears 'noisy' is that $\sigma_0$ was estimated voxel-wise using only 10 repeated acquisitions for each voxel. Only the lowest acquired flip angle data were used to estimate $\sigma_0$ since image subtraction revealed that this data contained the fewest artefacts. Note that the $SNR_{max}$ map here is not normalized by the number of phase-encoding steps, as this same factor multiplies the total repetition time of all SPGRs in $T_{acq}$, with both cancelling out. (d) To account for $B_1^+$ inhomogeneities, a separate $B_1^+$ map was acquired using AFI at a lower resolution, interpolated to the same resolution as the DESPOT1 data. All in-vivo data was reconstructed using a Roemer reconstruction (Roemer et al., 1990).

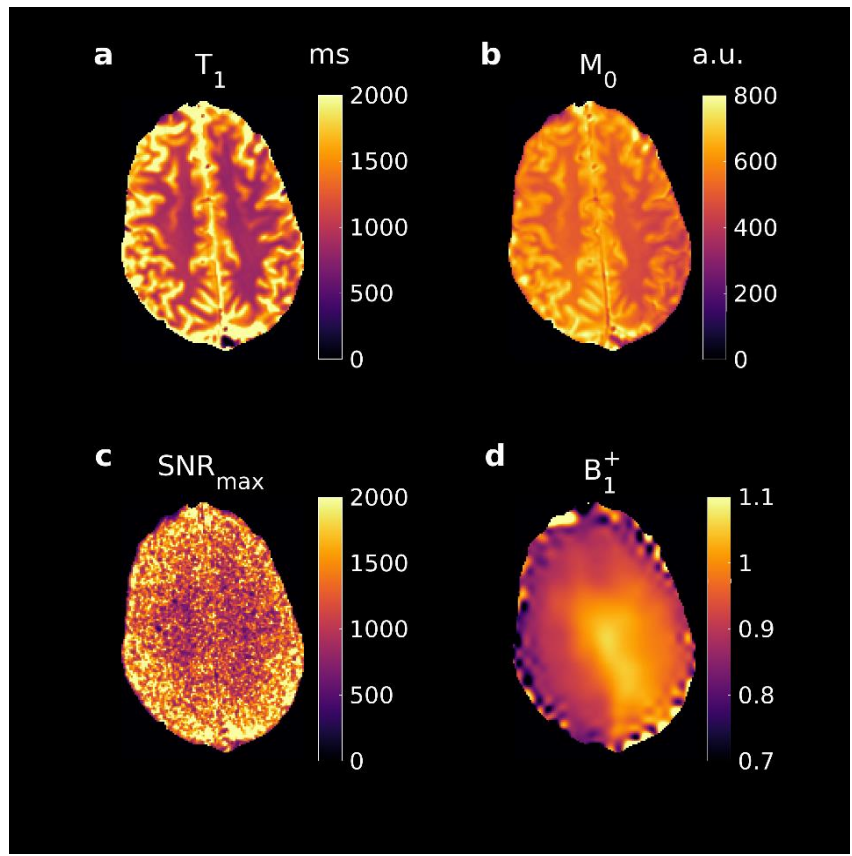